\documentstyle[buckow,epsfig]{article}
\input{epsf.tex}

\begin{document}
\def\titleline{
Topology in lattice QCD}
\def\authors{
A. Di Giacomo}
\def\addresses{
Dipartimento di Fisica and INFN, \\
2 Piazza Torricelli 56100 Pisa, Italy\\
digiacomo\@pi.infn.it}

\title{Topology in lattice QCD}
\author{ A. Di Giacomo\\
Dipartimento di Fisica and INFN, 2 Piazza Torricelli 56100 Pisa, Italy}
\def\abstracttext{
The status of topology on the lattice is reviewed. Recent results show that the
topological susceptibility $\chi$ can be unambigously determined. Different
methods, if properly implemented, give results consistent with 
each other. For $SU(3)$
the Witten -
Veneziano prediction is confirmed.
Preliminary results for full QCD are presented. The problem there is that the
usual hybrid montecarlo algorithm has severe difficulty to thermalize topology.
Possible ways out are under study.}
\makefront
\section{Introduction}
Topology plays a fundamental role in QCD. The key equation is the anomaly of the
$U_A(1)$ axial current\cite{1}
\begin{equation}
\partial^\mu J_\mu^5(x) = 2 N_f Q(x) \label{eq:1}\end{equation}
$J_\mu^5 = \sum_{f=1}^{N_f} \bar\psi_f \gamma^5 \gamma_\mu \psi_f$ is the
singlet axial current. $N_f$ is the number of light flavours, and
\begin{equation}
Q(x) = \frac{\displaystyle g^2}{\displaystyle 32\pi^2}
\sum_{a,\mu,\nu} G^a_{\mu\nu}{\tilde G}^{a\mu\nu}
\label{eq:2}\end{equation}
the topological charge density.
$\tilde G^a_{\mu\nu} = \frac{1}{2}\varepsilon_{\mu\nu\rho\sigma}
G^{a\rho\sigma}$
is the dual tensor field strength.
At the classical level $\partial^\mu J^5_\mu = 0$: the right hand side of  
eq.(\ref{eq:1}) comes from quantum effects, whence the name of anomaly.
$Q(x)$ is related to the Chern current $K_\mu(x)=
\frac{ g^2}{ 16\pi^2}
\varepsilon^{\mu\alpha\beta\gamma} A_\alpha^a\left(\partial_\beta A_\gamma^a -
\frac{g}{3} f^{abc} A_\beta^b A_\gamma^c\right)
$ as follows
\begin{equation}
\partial^\mu K_\mu = Q(x)\label{eq:3}\end{equation}
Eq.(\ref{eq:1}) can then be written
\begin{equation}
\partial^\mu (J^5_\mu - 2 N_f K_\mu) = 0 \label{eq:4a}\end{equation}
from which Ward identities are derived\cite{2}.

A number of physical consequences follow from eq.'s (\ref{eq:1}),
(\ref{eq:4a}).\par\noindent
{\bf 1)} The $U_A(1)$ problem. This problem goes back to Gellmann's free quark
model, from which the symmetries of hadron physics were abstracted. In that
model $U_A(1)$ is a symmetry, i.e. $\partial^\mu J^5_\mu=0$. This implies 
either
the existence of parity doublets in the hadron spectrum, if the symmetry is
Wigner, or, if it is Goldstone, Weinberg's inequality\cite{4} for the $\eta'$
mass
$
m_{\eta'} \leq \sqrt{3} m_\pi$.
Neither is true in nature.

In QCD $U_A(1)$ is not a symmetry, due to the anomaly (\ref{eq:1}) and in
principle there is no $U_A(1)$ problem. The anomaly can however explain the
high value of the $\eta'$ mass, as suggested by a
$1/N_c$ expansion. The idea is that in the limit $N_c\to
\infty$, with $g^2 N_c$ fixed the theory preserves all the main physical
features of QCD, like confinement\cite{5}. Non leading terms act as a perturbation on
this limiting model.
As $N_c\to\infty$ the anomaly disappears, being ${\cal O}(g^2)$ [eq.(2)], and
the $\eta'$ is the Goldstone boson of the spontaneously broken $U_A(1)$. The
effect of the anomaly is to shift the Goldstone pole from zero mass. This shift
can be computed by use of the Ward identities (\ref{eq:4a}) giving\cite{2}
\begin{equation}
\frac{\displaystyle 2 N_f}{\displaystyle f_\pi^2}\chi =
m_\eta^2 + m_{\eta'}^2 - 2 m_K^2
\label{eq:5}\end{equation}
$\chi = \left.\chi(q^2)\right|_{q=0}$ is the topological susceptibility of the
vacuum at $N_c=\infty$, or
\begin{equation}\hskip-18pt
\chi(q^2) = \int\! d^4x\, e^{iq x}\langle0| T\left(Q(x) Q(0)
\right)|0\rangle_{quenched}
\label{eq:6}\end{equation}
$N_c=\infty$ implies quenced approximation, fermion loops being ${\cal O}(g^2
N_f)$.

Eq.(\ref{eq:5}) predicts, within ${\cal O}(1/N_c)$ accuracy
\begin{equation}
\chi = (180\,{\rm MeV})^4 \label{eq:7}\end{equation}
To be definite a prescription must be given for the singularity in the
product $Q(x) Q(0)$ in eq.(\ref{eq:6}) as $x\to 0$.
The prescription which brings to eq.(\ref{eq:5})is\cite{2,6}
\begin{equation}
\hskip-10pt
\chi = \int d^4(x-y) \partial_\mu^x\partial_\nu^y\langle 0| T(K_\mu(x)
K_\nu(y))|0\rangle \label{eq:8}\end{equation}
By the prescription (\ref{eq:8}) $\delta$ like singularities in the product
$K_\mu(x) K_\nu(y)$ as $x\to y$ are eliminated after integration. In any 
regularization scheme only the multiplicative renormalization of $K_\mu$ is 
thus left.
Eq.(\ref{eq:5}) can be verified on the lattice, where $\chi_{quenched}$ can be
 computed. This computation provides a check of the $1/N_c$ expansion, which 
is a fundamental issue.\par\noindent
{\bf 2)} The behaviour of $\chi$ across the deconfining phase transition is 
relevant to understanding the structure of QCD vacuum\cite{7}.\par\noindent
{\bf 3)} The $U_A(1)$ Ward identities predict that in full QCD $\chi$ behaves 
in the chiral limit linearly in the  quark mass
\begin{equation}
\chi \simeq \frac{1}{N_f}\langle\sum_f m_f \bar\psi_f \psi_f\rangle
 \label{eq:9}\end{equation}
{\bf 4)} Another important quantity to determine on the lattice is $\chi'=
 \left.\frac{d}{d q^2}\chi(q^2)\right|_{q^2=0}$
This determination in full QCD is relevant to understand the spin content of 
the proton\cite{9}.\par\noindent
{\bf 5)} The spin content of the proton can be determined by use of Eq.(1), as
 will be discussed in detail in the following.

A lattice regulator $Q_L(x)$ of the operator $Q(x)$, eq.(\ref{eq:2}) is needed.
The regularized matrix elements are then determined by numerical simulations, 
and out of them  the continuum physical quantities can be extracted by proper 
renormalization.
\section{Defining $Q(x)$ on the lattice.}
According to the general rule any gauge invariant operator on the lattice 
$Q_L(x)$, such that in the formal limit $a\to 0$
\begin{equation}
Q_L(x) = Q(x) a^4 + {\cal O}(a^6)
\label{eq:2.1}\end{equation}
is a possible regulator of topological charge density.

A large arbitrariness by higher order terms exists in the choice of $Q_L$,
which 
can be used to improve the operator in
the sense which will be discussed below.
A prototype choice for $Q_L$ is
\begin{equation}\hskip-15pt
Q^{(0)}_L(x) =
-\frac{1}{32\pi^2}\sum_{\pm \mu\nu\rho\sigma} Tr\left[
\Pi_{(0)}^{\mu\nu}(x)\Pi_{(0)}^{\rho\sigma}(x)\right]
\label{eq:2.2}\end{equation}
where $\Pi_{(0)}^{\mu\nu}$ is the usual plaquette operator in the plane 
$\mu,\nu$. $Q^{(0)}_L$ obeys eq.(\ref{eq:2.1}), and  differs by ${\cal O}
(a^6)$ from any other choice. We will make use of a recursive improving of 
the operator (\ref{eq:2.2})\cite{10}. We shall define $Q^{(i)}_L(x)$
$(i = 1,2\ldots)$ 
by a formula similar to (\ref{eq:2.2}), with
 $\Pi_{(0)}^{\mu\nu}\Pi_{(0)}^{\rho\sigma}$ replaced by
$\Pi_{(i)}^{\mu\nu}\Pi_{(i)}^{\rho\sigma}$. For $i=1$ each link is replaced by
 a smoothed link, for $i=2$ each link of $\Pi_{(1)}$ is smoothed again
and so on\cite{10}.
An alternative definition is the so called geometrical charge\cite{11,12}, 
which again obeys the constraint eq.(\ref{eq:2.1}).

Any $Q_L$, in the limit $a\to 0$, will be a mixing of all the continuum 
operators with the same quantum numbers and with lower or equal dimension.
In quenched $QCD$ the only pseudoscalar with dimension $\leq 4$ is $Q(x)$ itself
and therefore\cite{13}
\begin{equation}
Q_L(x) \mathop\simeq_{a\to 0} Z\, Q(x)
\label{eq:2.4}\end{equation}
$Z$ can be determined by a non perturbative procedure known as 
``heating''\cite{14}. The idea is to measure $Q_L$ on a state with  a 
definite value of $Q$, which then determines $Z$. To do that an istanton is put
by hand on the lattice, and quantum fluctuations at a given value of 
$\beta = 2 N/g^2$ are produced numerically by the usual updating algorithm. A
plateau after a few heating steps signals thermalization of these 
fluctuations, while the topological content of the configuration 
takes a much longer 
time to be changed, as can be directly tested: the values of $Z$ for 
different
operators $Q_L$ 
at different $\beta$'s are shown in Table~1.
\vskip0.1in\par\noindent\centerline{
\begin{tabular}{|c|c|c|c|c|c|c|}
\hline
 & Op & $\chi_L\cdot 10^{-5}$ & $ M\cdot 10^{-5}$ & $M/\chi_L$ & $Z$ &
$\chi^{1/4}$~MeV\\ \hline
$ N_c = 3$ 
              & $Q_0$ & 2.72(6) & 2.50(15) & .9 & .12(4) & 167(36)\\ 
$\beta = 5.9$ & $Q_1$ & 2.48(5) & .88(6)   & .35& .36(2) & 178(6) \\
$a^{-1} = 1.74(4)\,{\rm GeV}$
              & $Q_2$ & 3.51(7) & .71(6)   & .2& .48(2) 	& 175(6) \\ \hline
$ N_c = 3$ 
              & $Q_0$ & 2.14(4) & 1.97(10) & .9 & .18(4) & 196(40)\\ 
$\beta = 6.1$ & $Q_1$ & 1.12(2) & .47(3)   & .42& .41(2) & 178(6) \\
$a^{-1} = 2.17(5)\,{\rm GeV}$
              & $Q_2$ & 1.39(2) & .33(2)   & .24& .54(2) 	& 175(6) \\ \hline
$ N_c = 2$ 
              & $Q_0$  & 3.38(8) & 2.93(7) & .87 & .18(6)    & 208(23)\\ 
$\beta = 2.44$ & $Q_1$ & 3.26(8) & 1.45(4) & .44 & .405(13)  & 197(8) \\
$a^{-1} = 1.92\,{\rm GeV}$
              & $Q_2$ & 4.92(14) & 1.34(5) & .27 & .568(16) 	& 197(8) \\ 
       & $Q_{Geo}$ &$69.7\pm1.1$& 52(2)   & .75 & .918(57) 	& 231(13)\\ \hline
$ N_c = 2$ 
              & $Q_0$  & 2.32(5) &2.20(3)  & .95 & .240(26)    & 197(27)\\ 
$\beta = 2.57$ & $Q_1$ & 1.01(5) & .44(2)  & .45 & .507(9)  & 200(8) \\
$a^{-1} = 2.91(9)\,{\rm GeV}$
              & $Q_2$ & 1.16(6)  & .117(5) & .16 & .675(8) 	& 198(7) \\ 
       & $Q_{Geo}$ & 16.6(3)& 13.26(23)    & .80 & .937(26) 	& 228(12)\\ \hline
\end{tabular}
}\vskip0.05in
{\bf Table 1.} $\chi_L, M, \chi, Z, a$ for quenched $SU(3)$ and $SU(2)$
at various $\beta$.
\section{Measuring $\chi$}
The lattice susceptibility is defined as
\begin{equation}
\chi_L = \sum_x \langle Q_L(x) Q_L(0)\rangle = 
\frac{\displaystyle \langle Q_L^2\rangle}{\displaystyle V}
\label{eq:3.1}\end{equation}
$\chi_L$ is a positive quantity by definition.
However in the euclidean region, for $x\neq 0$, $ \langle Q(x) Q(0)\rangle$
is a negative quantity, by reflection positivity, $Q$ being odd under time 
reflection.
This can be checked on the lattice and the result is shown in fig.1. If the 
operator is smeared over  a region of size $s$, one expects 
$ \langle Q(x) Q(0)\rangle$ to be negative at distances $|x|\geq s$. 
The peak at $x=0$ is essential to make $\chi$
positive, and hence the prescription for the product at $x=0$ is essential.
In general, as a consequence of operator product expansion around $x=0$, 
$\chi_L$ will mix with the continuum operators with the same quantum numbers
and lower or equal dimension\cite{15}
\begin{equation}
\chi_L = Z^2 \chi a^4 + M(\beta) + {\cal O}(a^6)
\label{eq:3.2}\end{equation}
with
\begin{equation}
M(\beta) =
\overline Z(\beta) a^4\langle \frac{\beta(g)}{g} G^a_{\mu\nu} G^a_{\mu\nu}
\rangle
+\overline{\overline Z}(\beta) a^4
\langle m\bar\psi(x)\psi(x)\rangle + P(\beta)\langle I\rangle
\label{eq:3.3}
\end{equation}
In the quenched approximation the second term in eq.(\ref{eq:3.3}) will be
absent.

The first term in  eq.(\ref{eq:3.2}) corresponds to the prescription
(\ref{eq:8}) and renormalizes multiplicatively. 
Since
$ \int d^4x\,\partial_\mu K^\mu =Q$
the prescription (\ref{eq:8}) 
implies that $\chi$ must be zero on the sector $Q=0$.
$M(\beta)$ can be obtained by measuring $\chi_L$ on that sector. This is 
done by the same heating technique\cite{14} used to 
determine $Z$: a sample of 
configurations belonging to the $Q=0$ sector is produced by heating  the flat 
(zero field)  configuration,  in such a way that the topological charge is not 
changed.
From eq.(\ref{eq:3.2}) then
\begin{equation}
\chi = \frac{\chi_L - M(\beta)}{a^4(\beta) Z^2(\beta)}
\label{eq:3.4}\end{equation}
All the quantities in the r.h.s. of eq.(\ref{eq:3.4}) depend on the choice of 
the operator, and/or on the choice of the action, but the result must be 
independent of them. This appears from Table~1, and from fig.2,3.

A good operator is such that $|M(\beta)|/|\chi_L| \ll 1$ so that most of the 
observed signal is physcal. $Z\simeq 1$ is also desirable.
The quality of different $Q_L$'s can be appreciated from Table~1.
 
The behaviour of $\chi$ at 
deconfinement is shown in fig.2 \cite{16}, where $SU(2)$ and $SU(3)$ can be 
directly  compared.

The existence on lattice of a nontrivial $Z$ was first realized in ref.13.
Eq.(\ref{eq:3.4}) was first introduced in ref.11, where, however, the 
renormalization constants were determined by perturbation theory.

The heating technique\cite{14} finally allowed a non perturbative determination
 of them.
\vskip0.1in\par\noindent
\begin{minipage}{0.5\textwidth}
\epsfxsize = 0.5\textwidth
{\centerline{
\rotatebox{-90}{
{\epsfbox{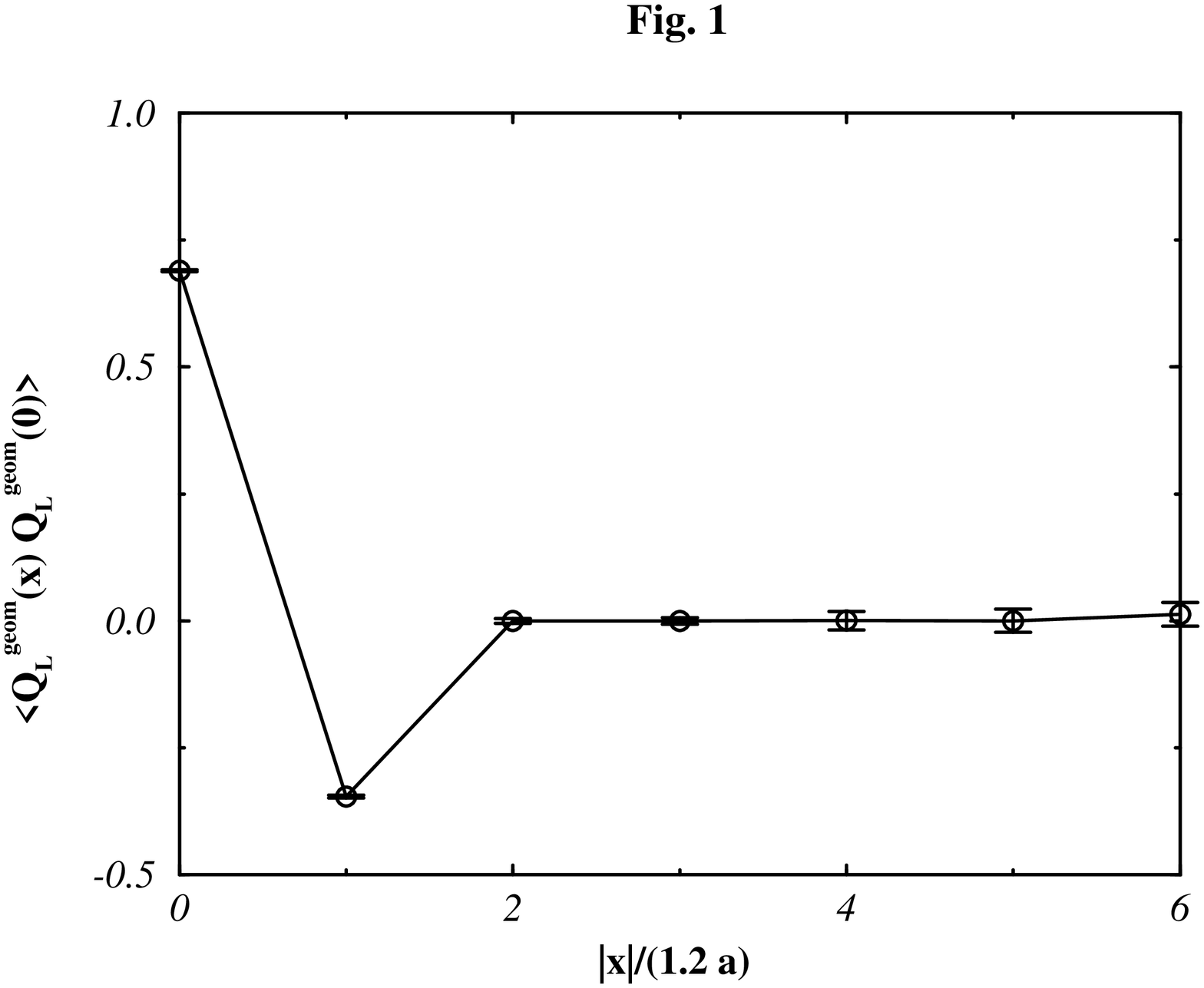}}
}
}}
\vskip0.1in\par\noindent
{\bf fig.1} Correlation $\langle Q(x) Q(0)\rangle$ for the
geometric \\ 
topological charge density.
\end{minipage}
\begin{minipage}{0.5\textwidth}
\epsfxsize = 0.5\textwidth
{\centerline{
\rotatebox{-90}{
{\epsfbox{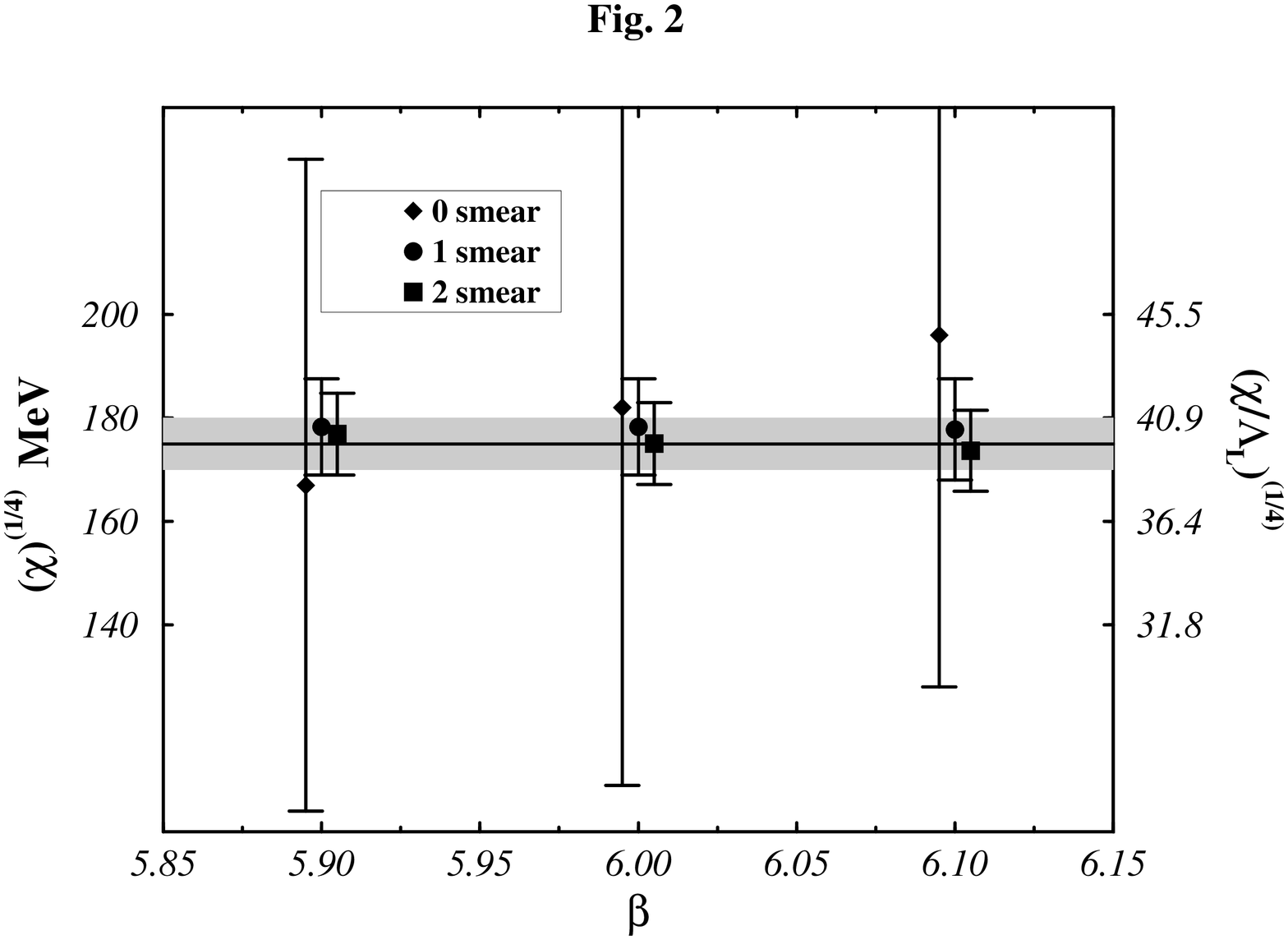}}
}
}}
\vskip0.1in\par\noindent
{\bf fig.2} $\chi$ for quenched $SU(3)$. Diamonds, circles and squares corrspond
to 0,1 and 2 - smeared operators.
\end{minipage}
\vskip0.1in\par\noindent
\begin{minipage}{0.5\textwidth}
\epsfxsize = 0.5\textwidth
{\centerline{
\rotatebox{-90}{
{\epsfbox{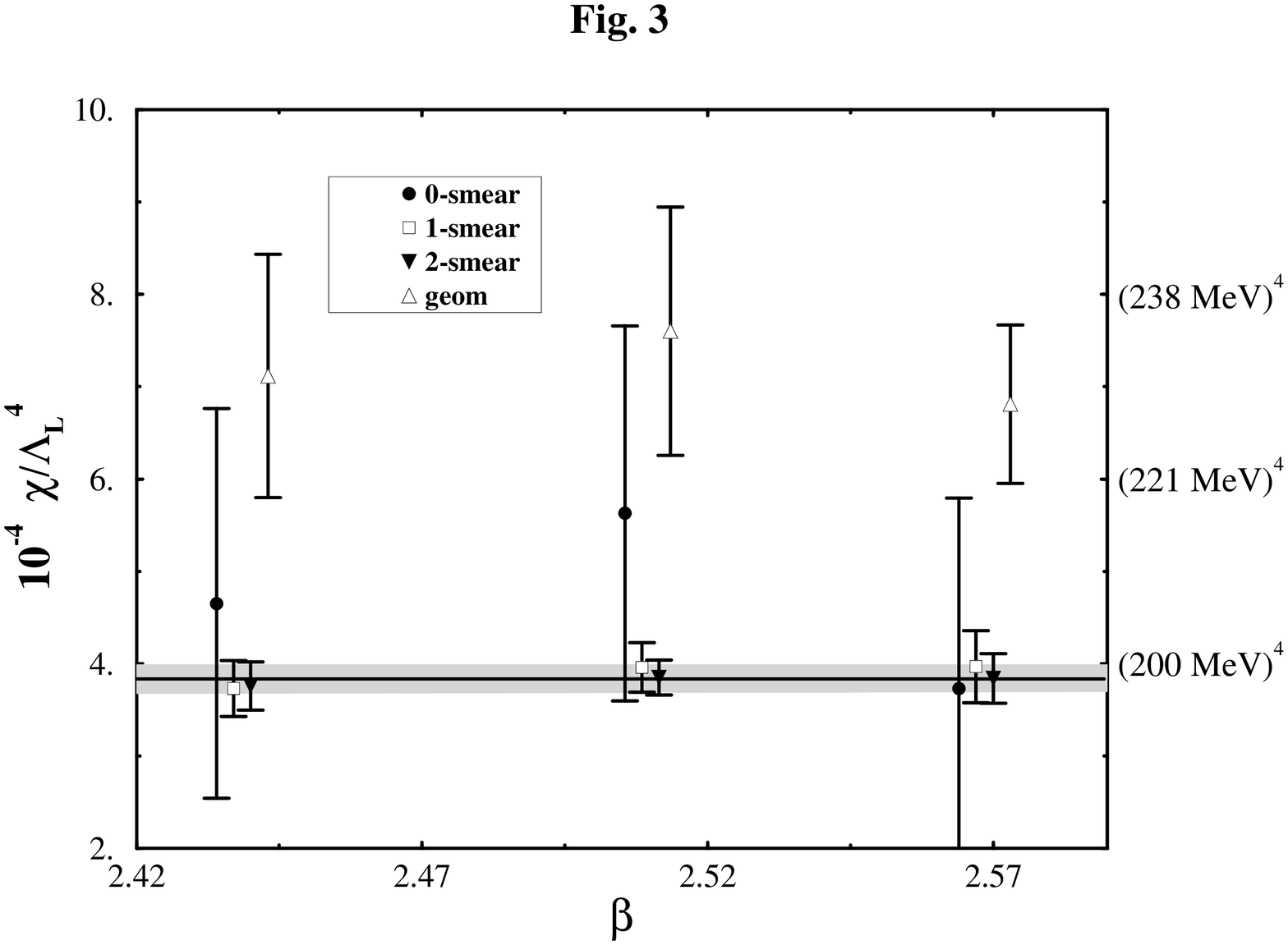}}
}
}}
\vskip0.1in\par\noindent
{\bf fig.3} $\chi$ for quenched $SU(2)$. 
\end{minipage}
\begin{minipage}{0.5\textwidth}
\epsfxsize = 0.5\textwidth
{\centerline{
\rotatebox{-90}{
{\epsfbox{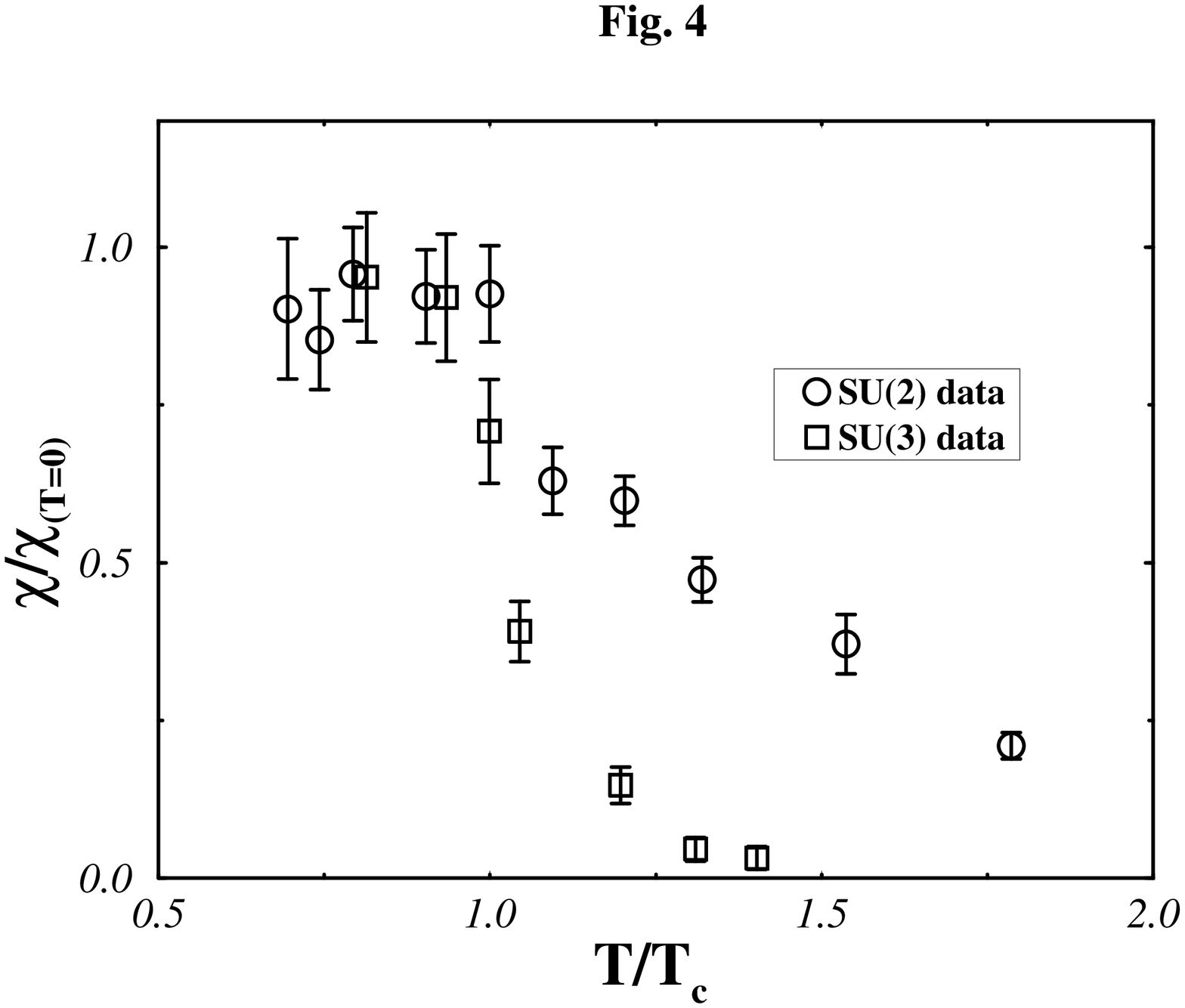}}
}
}}
\vskip0.1in\par\noindent
{\bf fig.4} $\chi$ across the deconfinement transition for $SU(2)$ (circles)
and $SU(3)$ (squares).
\end{minipage}

\section{Full QCD$^{17}$.}
We have used the same technique to determine $\chi$ in full QCD with staggered
fermions at $m_Q = 0.01$ and $m_Q=0.02$ at $\beta = 5.35$. The lattice was
$16^3\times24$. At this $\beta$ value $a = 0.115(2)$~fm.

We do not have enough precision to test the dependence on $m_Q$
[eq.(\ref{eq:9})]. At $a m = 0.01$ we get
$
\chi = (110 \pm 6 \pm 2)^4\,{\rm MeV}^4
$
The first error is obtained by jake-knife technique. The second comes from the
error in the determination of the scale. The expected
value is $\chi \simeq(109\,{\rm MeV})^4$.

As we will soon discuss these
errors are higly underestimated.

A similar analysis can be done for $\chi'$. On general grounds
\begin{equation}
\chi' = Z^2 \chi' a^2 + M'(\beta) + {\cal O}(a^4)
\qquad
M'(\beta) = P'(\beta) \langle I \rangle
\label{eq:4.2}\end{equation}
$\chi'$ only mixes to the identity operator, since no other gauge invariant
operator of dimension $\leq 2$ exists.

The technique used to determine $\chi'$ is to improve the operator enough, so
that $M'(\beta)$ will be negligeable as compared to the first term. $Z$ is
known from the analysis of $\chi$.

A preliminary result is
\begin{equation}
\chi' = (258\pm 100)\,{\rm MeV}^2
\qquad \sqrt{\chi'} = (16\pm 3)\,{\rm MeV}
\label{eq:4.3}\end{equation}
to be compared with the value computed from SVZ rules\cite{9}
$ \sqrt{\chi'} = (22.3\pm 4.8)\,{\rm MeV}$
Here the error coming from the normalization is much smaller than the
statistical error.

Again here the error is underestimated. The reason can be 
in
the history of the topological charge along the updating process is shown. The
hybrid montecarlo algorithm is very slow in thermalizing topology\cite{18}.
The sampling is very bad and corresponds in fact to a much smaller number of 
independent configurations.

The same incovenience affects the determination of the spin content of the
proton which will be discussed in the next section. The real error is then
larger than the one estimated on the present ensemble of configurations.
\section{The spin content of the proton.}
The matrix element of the singlet axial current between proton states can be
written as
\begin{equation}
\langle \vec{p}' s'| J_\mu^5(0) |\vec p s\rangle =
 \bar u( \vec{p}' s')\left\{
G_1(k^2) \gamma^5 \gamma_\mu + G_2(k^2) \gamma^5 k_\mu\right\}
u(\vec p s)
\label{eq:5.1}\end{equation}
$k = p-p'$ is the momentum transfer.
$G_1(0) \equiv \Delta \Sigma$ is related to the integral of the spin dependent
structure function $g_1(x,q^2)$ of deep inelastic scattering of leptons off
protons. In naive parton picture
$\Delta \Sigma = \Delta u + \Delta d + \Delta s$ 
is the fraction of theproton spin carried by the quarks and is expected to be
$\Delta \Sigma = 0.7$. Experiments\cite{19} show that it is much 
smaller,
$\Delta \Sigma = 0.21\pm0.10$, and this fact is usually referred as spin 
crisis.
By use of the anomaly equation
\[
\langle \vec{p}' s'| Q(0) |\vec p s\rangle =
\frac{1}{2 N_f}\langle \vec{p}' s'| \partial^\mu J_\mu^5(0) |\vec p s\rangle 
 = \frac{m_N}{N_f} A(k^2) 
\bar u( \vec{p}' s') i\,\gamma^5
u(\vec p s)
\]
$A(k^2) = G_1(k^2) + (k^2/m_N) G_2(k^2)$ can be extracted from a measurement of
the matrix element in the left hand side of eq.(\ref{eq:5.1}).

In quenched approximation $G_2(k^2)$ has a pole at $k^2=0$\cite{22}. With
dynamical quarks this is ot the case and, for sufficiently small $k^2$ $G_1(0)$
or $\Delta \Sigma$\cite{23} can be determined.

Our preliminary result is
$
\Delta \Sigma = 0.05\pm 0.05$
Again the error is underestimated due to the bad sampling of topological charge
discussed above.
\section{Discussion.}
Our conclusions are the following.

A reliable determination of the topological susceptibility $\chi$ is
possible on the lattice. All existing definitions of lattice topological charge 
density give results consistent with each other if proper renormalizations are
performed.
The value of $\chi$ for $SU(2)$ is larger than the prediction\cite{2}
but the value for $SU(3)$ is consistent with it. The value of $m_{\eta'}$
is well explained by the anomaly.
Above the deconfining transition $\chi$ drops to zero, 
more rapidly for $SU(3)$ 
than for $SU(2)$\cite{16}.
In full QCD preliminary results are in agreement with expectations for $\chi$
and $\chi'$ and for the spin content of the proton.
However a better sampling of topological sectors is needed. In fact the 
incapability  of the hybrid montecarlo to thermalize topology could affect not 
only the measurement of quantities directly related to it, like the one we 
considered, but any other measurement on the lattice.
Work is in progress to overcome this difficulty.

I am grateful to my collaborators B. Alles, G. Boyd, M. D'Elia, E. Meggiolaro,
H. Panagopoulos, with whom most of the results presented where obtained.

\end{document}